\pdfoutput=1

\documentclass[aps,prd,twocolumn,superscriptaddress,showpacs]{revtex4}
\usepackage{graphics}
\usepackage{epstopdf}
\usepackage{graphicx}
\usepackage{dcolumn}
\usepackage{bm}
\usepackage{amsmath}

\newcommand{\be}{\begin{equation}}
\newcommand{\ee}{\end{equation}}
\newcommand{\ba}{\begin{eqnarray}}
\newcommand{\ea}{\end{eqnarray}}
\newcommand{\bfig}{\begin{figure}[t]\begin{centering}}
\newcommand{\efig}{\end{centering}\end{figure}}

\def\dm03{\hbox{$\Delta m^2_{03}$}}

\begin{document}


\title{Searches  for sterile neutrinos with IceCube DeepCore}

\author{Soebur Razzaque}
\email{srazzaqu@gmu.edu}
\affiliation{School of Physics,  Astronomy  and Computational Sciences, George Mason University, Fairfax, VA 22030, USA}
\affiliation{
Resident at the Space Science Division, US Naval Research Laboratory, Washington, DC 20375, USA}
\author{A.\ Yu.\ Smirnov}
\email{smirnov@ictp.it}   
\affiliation{The Abdus Salam International Centre for Theoretical Physics, I-34100 Trieste, Italy}

\date{\today}
 
\begin{abstract}
We show that study of the atmospheric neutrinos in the 10--100 GeV
energy range by DeepCore sub-array of the IceCube Neutrino Observatory
can substantially constrain the mixing of sterile neutrinos of mass
$\sim 1$ eV with active neutrinos.  In the scheme with one sterile
neutrino we calculate $\nu_\mu-$ and $\bar{\nu}_\mu-$ oscillation
probabilities as well as zenith angle distributions of $\nu_\mu^{CC}$
(charge current) events in different energy intervals in DeepCore.
The distributions depend on the mass hierarchy of active
neutrinos. Therefore, in principle, the hierarchy can be identified,
if $\nu_s$ exists.  After a few years of exposure the DeepCore data
will allow to exclude the mixing $|U_{\mu 4}|^2 \geq 0.02$ indicated
by the LSND/MiniBooNE results. Combination of the DeepCore and high
energy IceCube data will further improve sensitivity to $\nu_s$ mixing
parameters.
\end{abstract}                            
 
\maketitle

\section{Introduction}

Mixing of sterile neutrinos of eV mass scale with active neutrinos, as
is indicated by the LSND/MiniBooNE results~\cite{lsnd,miniboone},
significantly affects the atmospheric neutrino fluxes
~\cite{nunokawa,choubey,Razzaque:2011ab,Barger:2011rc,Halzen:2011yq}.
The primary effect arises in the TeV energy range due to the MSW
resonant enhancement of ${\bar \nu}_\mu - {\bar \nu}_s$ oscillations
while neutrinos pass through the matter of the Earth.  As a result,
the ${\bar \nu}_\mu-$flux, and consequently, number of $\nu_\mu-$
events at the detectors deplete.  The oscillations lead to distortion
of the zenith angle and energy distributions of events.

In this paper we will show that the atmospheric neutrino fluxes at low
energies, $E < 0.5$ TeV, are also strongly affected by the $\nu_s$
oscillations in spite of the fact that the resonance enhancement
occurs in the TeV range. Therefore studies of the low energy fluxes
can substantially contribute to searches for sterile neutrinos.  With
a densely spaced array of phototubes in the middle of the IceCube
array, DeepCore sub-array~\cite{deepcore}, \cite{Ha:2012ww}
significantly enhances IceCube sensitivity to $(10 - 100)$~GeV
neutrinos.  This makes possible to study the atmospheric neutrino
oscillations with DeepCore at low
energies~\cite{Mena:2008rh,FernandezMartinez:2010am,Abbasi:2011zz,Fargion:2011tx}.

As it was realized in Ref.~\cite{Razzaque:2011ab}, the $\nu_s$ effects
depend not only on $U_{\mu 4}$ relevant for the LSND/MiniBooNE results
but also on the $\nu_\tau$ admixture $U_{\tau 4}$.  There are two
special cases: (i) $U_{\tau 4} = 0$, which corresponds to the so
called flavor-mixing scheme, and (ii) $U_{\tau 4} = U_{\mu 4}$ which
determines the mass-mixing scheme for maximal 2-3 mixing. In the
latter case $\nu_s$ mixes with the active neutrinos through the mass
eigenstate $\nu_3$.  At energies below $0.5$~TeV in the $\nu_s$
mass-mixing scheme the $\nu_s$ effects are strong, whereas in the
$\nu_s$ flavor-mixing scheme the oscillation probabilities are
unaffected~\cite{Razzaque:2011ab}.  In this paper for illustration we
consider a scheme with one sterile neutrino, as was favored by recent
MINOS data~\cite{minos2011,Sousa:2011rw} and global
fit~\cite{Giunti:2011hn} and focus on the $\nu_s$ mass-mixing scheme.
We show that the mixing angle with the active neutrinos can be
significantly constrained by $(10 - 100)$~GeV atmospheric neutrino
data collected in DeepCore~\cite{deepcore}.

We present oscillation probabilities in the $\nu_s$ mass-mixing scheme
at low energies in Sec.~II.  In Sec.~III we calculate the
$\nu_\mu^{CC}$ event rate in DeepCore produced by atmospheric
neutrinos. We study the zenith and energy distributions of these
events.  We estimate the sensitivity of DeepCore to the $\nu_s$ mixing
with active neutrinos in Sec.~IV. Sec. V contains discussion and
conclusions.

\section{Oscillation Probabilities}

At high energies to a good approximation one can consider mixing of
the three flavor states $\nu_f^T\equiv (\nu_s, \nu_\tau, \nu_\mu)$ in
the mass eigenstates $\nu_{mass}^T \equiv (\nu_4, \nu_3, \nu_2)$,
neglecting the mixing of $\nu_1$~\cite{Razzaque:2011ab}.  In the
$\nu_s$ mass-mixing scheme, the mixing matrix $U_f$, defined as $\nu_f
= U_f\nu_{mass}$, depends on the angle $\theta_{23}$ and new
``sterile'' angle $\alpha$, so that in $\nu_4$
$$
U_{s4} = \cos\alpha,~ U_{\mu 4} = \sin\alpha \sin\theta_{23}, ~
U_{\tau 4} = -\sin\alpha \cos\theta_{23}.
$$  

For $\alpha = 0$, when the state $\nu_4$ decouples and the
3$\nu-$evolution reduces to 2$\nu-$evolution, the $\nu_\mu$ survival
probability is given by
\be
P_{\mu\mu} = 1- \sin^2 2\theta_{23} \sin^2 \frac{\Delta m_{32}^2 x}{4E}. 
\label{prob2nu}
\ee 
Here $x= 2R_\oplus |\cos \theta_z|$ is the length of neutrino
trajectory characterized by the zenith angle $\theta_z$, and
$R_\oplus$ is the radius of the Earth.  In our numerical estimations
we use $\Delta m_{32}^2 = (2.3 - 2.5) \cdot 10^{-3}$~eV$^{2}$ and
$\sin^22\theta_{23} = 1$.

According to Eq.~(\ref{prob2nu}) the first oscillation minimum (dip)
is at energy
\be
E_{\rm min}^0 = \frac{1}{\pi} \Delta m_{32}^2 R_\oplus  \cos \theta_z.  
\label{oscdip1}
\ee
For neutrinos moving along the Earth's diameter, $\theta_z = \pi$, we
obtain $E_{\rm min}^0 \approx 25.7$ ~GeV.

If $\alpha \neq 0$, the oscillation probability changes substantially
due to matter effect.  For $E > 15$ GeV to a good approximation the
effect is described by oscillations in uniform medium with the average
density along the neutrino trajectory. In turn, for constant density
and for energies below 0.5 TeV the probability is given
by~\cite{Razzaque:2011ab}
\ba
P_{\mu\mu} &=& 1- \cos^2\alpha \sin^22\theta_{23} 
\nonumber \\ && \times 
\sin^2 \left[ \left(\frac{\Delta m_{32}^2}{2E} 
\pm |V_\mu (\theta_z)|\sin^2\alpha  \right) 
R_\oplus \cos \theta_z 
\right]
\nonumber \\ &&
- 0.5 \sin^2 2\alpha \sin^4\theta_{23} -0.5 \sin^2\alpha \sin^22\theta_{23},
\label{probnus}
\ea
after averaging over fast oscillations driven by the mass squared
difference $\Delta m_{43}^2 \sim 1$~eV$^2$. Here $V_\mu (\theta_z)$ is
the average matter potential: $V_\mu(\theta_z) = -G_F
n_n(\theta_z)/\sqrt{2}$, and $n_n(\theta_z)$ is the average number
density of neutrons along the $\theta_z$ trajectory.  The lower
``$-$'' (the upper ``$+$'') sign in front of $|V_\mu|$ in
Eq.~(\ref{probnus}) corresponds to antineutrinos (neutrinos).  Thus,
in the first approximation the $\nu_s$ effect is reduced to appearance
of an additional contribution to the oscillation phase.  This
contribution has different signs in the $\nu-$ and
${\bar \nu}-$channels as well as for normal (NH) and inverted (IH)
mass hierarchies of the active neutrinos ({\it i.e.} for the positive
and negative signs of $\Delta m_{32}^2$).

According to Eq.~(\ref{probnus}) the energy of the first oscillation
dip changes due to $\nu_s$ mixing as
\ba
E_{\rm min} & \approx & \frac{|\Delta m_{32}^2| }
{\frac{\pi}{R_\oplus \cos \theta_z}  \mp 2|V_\mu (\cos \theta_z)| \sin^2\alpha}
\nonumber \\ &  = &
\frac{E_{min}^0}{1\mp \frac{2}{\pi} |V_\mu (\cos \theta_z)| 
R_{\oplus} \sin^2\alpha \cos \theta_z},  
\label{oscdip2}
\ea
where the lower sign in denominator corresponds to antineutrinos for
the normal mass hierarchy and to neutrinos for the inverted mass
hierarchy.  For $\sin^2 \alpha = 0.04$, fixed $\Delta m^2_{32}$, NH
and $\cos \theta_z = -1$ (an average density $\sim 8.4$~g~cm$^{-3}$)
the oscillation dips are at $29.0$~GeV and $22.4$~GeV for neutrinos
and antineutrinos, respectively.  This $\nu_s$ induced shift of the
$\nu_\mu - \nu_\mu$ oscillation dip can be used to constrain the
mixing angle $\alpha$.

We computed the oscillation probabilities as functions of neutrino
energy and zenith angle by numerically solving the $3\nu-$evolution
equation~\cite{Razzaque:2011ab} with the density profile given by the
Preliminary Reference Earth Model~\cite{prem}. The features of the
obtained results can be easily understood with Eqs. (\ref{prob2nu}) to
(\ref{oscdip2}).

\bfig
\includegraphics[trim=0.25in 0.in 0.9in 0.in, clip, width=3.35in]{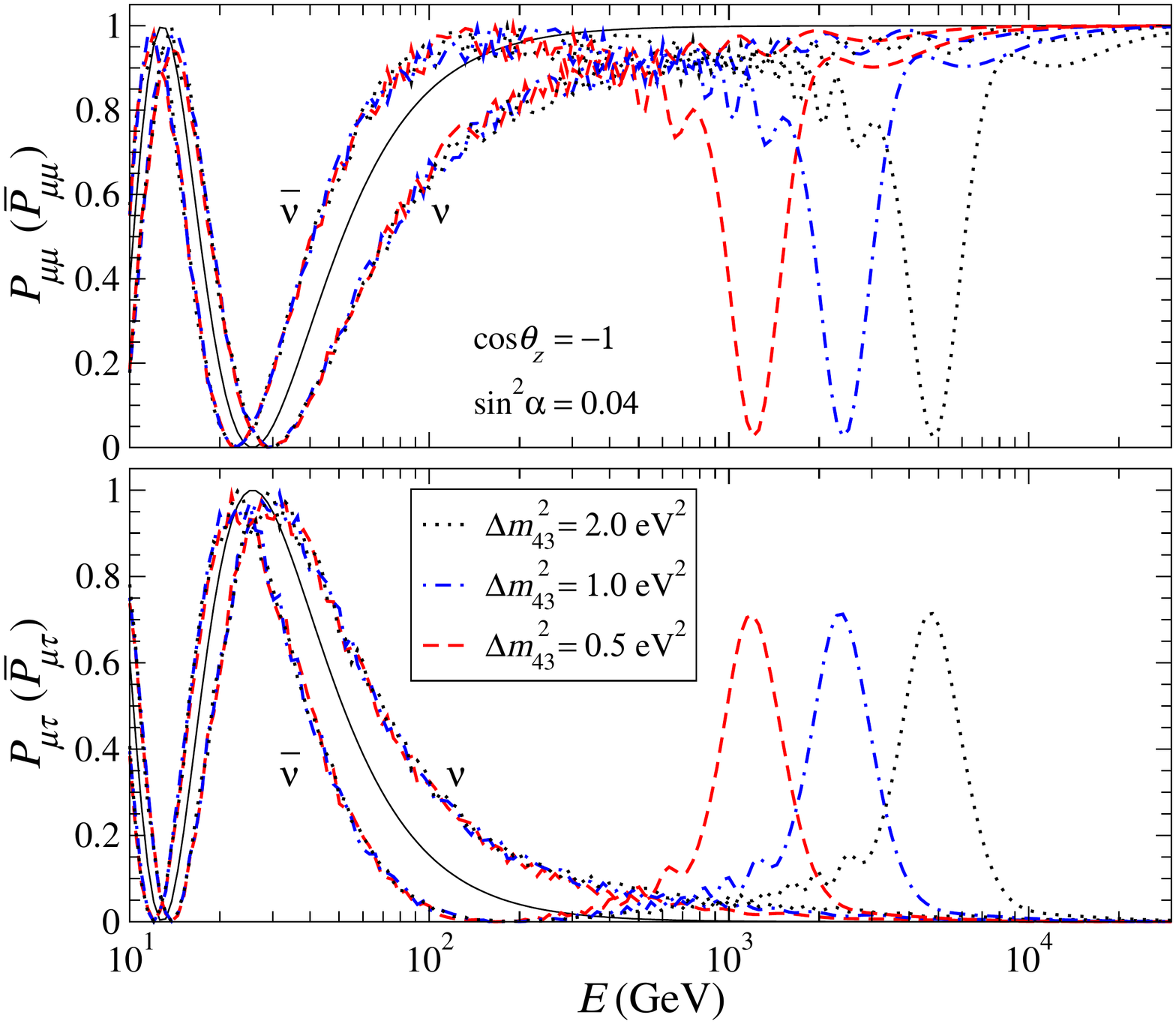}
\caption{
Oscillation probabilities $\nu_\mu \to \nu_\mu$ ({\em top panel}) and
$\nu_\mu \to \nu_\tau$ ({\em bottom panel}) for different values of
$\Delta m^2_{43}$. The solid (broken) lines correspond to the
probabilities without (with) $\nu_s$ mixing. The normal mass hierarchy
is assumed.
}
\label{fig:prob}
\efig

\bfig
\includegraphics[trim=0.25in 0.in 0.9in 0.in, clip, width=3.35in]{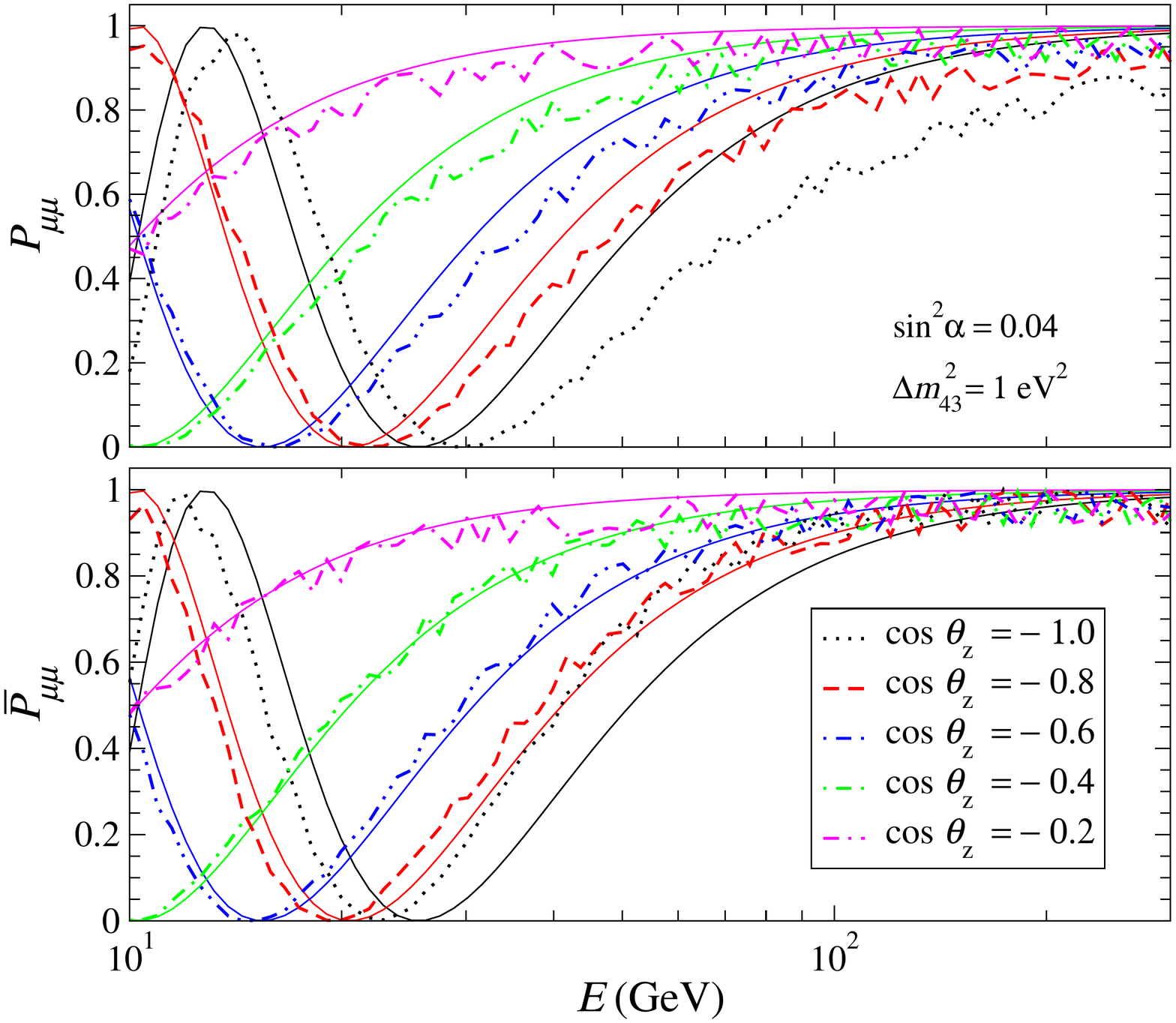}
\caption{
The survival probabilities of $\nu_\mu$ ({\em top panel}) and
${\bar \nu}_\mu$ ({\em bottom panel}) for different values of the
zenith angle.  The solid (broken) lines correspond to the
probabilities without (with) $\nu_s$ mixing. The normal mass hierarchy
is assumed.
}
\label{fig:prob_mm}
\efig

\bfig
\includegraphics[trim=0.25in 0.in 0.9in 0.in, clip, width=3.35in]{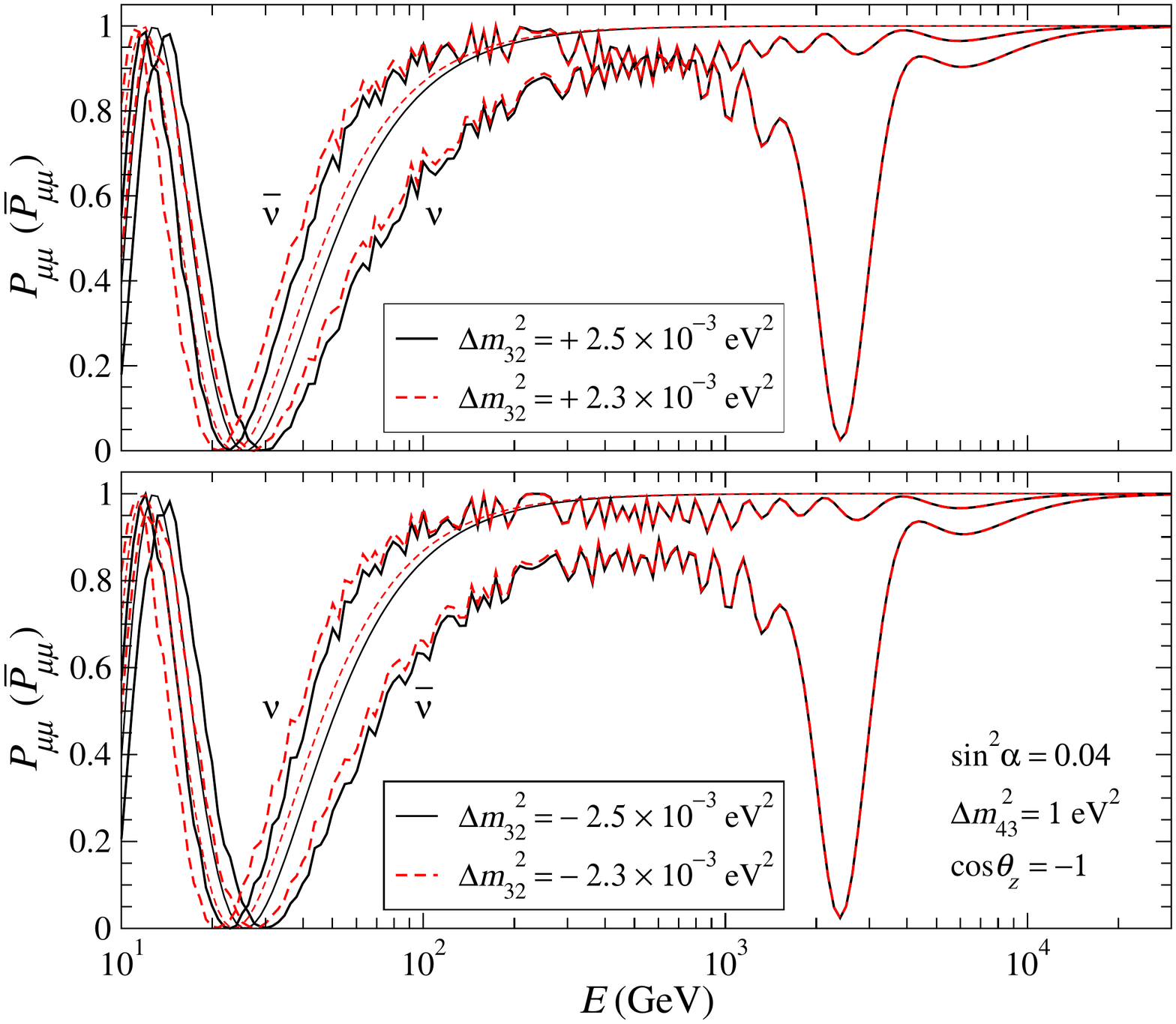}
\caption{
The $\nu_\mu-$ and ${\bar \nu}_\mu-$ survival probabilities for two
different absolute values of $\Delta m^2_{32}$.  {\it Top panel}:
normal mass hierarchy, $\Delta m^2_{32}> 0$; {\it bottom panel}:
inverted mass hierarchy, $\Delta m^2_{32} < 0$.  Thin lines are for
probabilities without $\nu_s$ mixing.
}
\label{fig:prob_dms}
\efig

Figure~\ref{fig:prob} shows the $\nu_\mu \to \nu_\mu$ (top panel) and
$\nu_\mu \to \nu_\tau$ (bottom panel) oscillation probabilities as
functions of neutrino energy for different $\Delta m_{43}^2$ but fixed
zenith angle $\theta_z = \pi$.  The energies of the oscillation dips
are in good agreement with the approximate values from
Eqs.~(\ref{oscdip1}) and (\ref{oscdip2}).  According to
Fig. ~\ref{fig:prob} the energy of resonant oscillation dip in the
${\bar \nu}-$channel is proportional to $\Delta m_{43}^2$.  In
contrast, at energies $\lesssim 100$~GeV the probabilities depend on
$\Delta m_{43}^2$ very weakly, which can be seen explicitly in
Eq. (\ref{probnus}). From the bottom panels, we find that at low
energies $\nu_\mu$ mainly transforms to $\nu_\tau$, whereas at high
energies $\nu_\mu \rightarrow \nu_s$ transition is significant.

Figure~\ref{fig:prob_mm} shows the probabilities $P_{\mu\mu}$ and
${\bar P}_{\mu\mu}$ as functions of neutrino energy for different
values of $\cos\theta_z$, but fixed $\sin^2\alpha$ and $\Delta
m_{43}^2$ and for NH.  With decrease of $|\cos \theta_z|$ the length
of the neutrino trajectory decreases and $E_{\rm min}$ becomes
smaller.  Furthermore, the difference of probabilities with and
without $\nu_s$ decreases.

In Figure~\ref{fig:prob_dms} we show dependence of the survival
probabilities on the absolute value of $\Delta m^2_{32}$ and on its
sign, {\it i.e.}  on mass hierarchy of active neutrinos.  A variation
of $\Delta m_{32}^2$ also shifts the dip.  However, due to dependence
of $V$ on $\theta_z$ and energy dependence of the usual
($\propto \Delta m_{32}^2$) term in Eq. (\ref{probnus}) the effects of
$\Delta m_{32}^2$ variations and of sterile neutrino are
different. Thus, studies of the energy and zenith angle dependence of
the $\nu_\mu$ event rates will allow to disentangle the two effects.
Change of the mass hierarchy leads to interchange of $\nu$ and
$\bar{\nu}$ probabilities at low energies. As a result, for IH the
$\bar{\nu}$ probability is below the $\nu$ probability in whole energy
range, and there is no intersection of the two at 0.5 - 0.7 TeV.  As
follows from Figure~\ref{fig:prob_dms}, the effect of $\Delta
m_{32}^2$ uncertainty (within 68\% CL region allowed by the MINOS
results~\cite{Adamson:2012rm}) on the probabilities, is relatively
small.  However, it becomes significant when events from $\nu$ and
$\bar{\nu}$ interactions sum up.

\section{Event rates in DeepCore}

We calculate the $\nu_\mu^{CC}-$ (charged current) event rate in
DeepCore.  The $\nu_\mu-$flux at the detector equals $\Phi_\mu
= \Phi_\mu^0 P_{\mu\mu} + \Phi_e^0 P_{e\mu} \approx \Phi_\mu^0
P_{\mu\mu}$, where the original atmospheric $\nu_\mu-$flux,
$\Phi_\mu^0$, is larger than the original $\nu_e-$flux, $\Phi_e^0$, by
a factor $> 10$ in the energy range under consideration.  Furthermore,
$P_{e\mu} \ll 1$ and $\nu_e$ mostly converts into
$\nu_s$~\cite{Razzaque:2011ab}.  So, we will neglect the effect of
$\nu_e \rightarrow \nu_\mu$ transition.  However, we do take into
account contributions from the tau leptons, created by the $\nu_\tau
N$ interactions. Tau leptons decay into muons with branching ratio
$\epsilon \sim 0.18$ being recorded as $\nu_\mu^{CC}-$events.  The
$\nu_\tau-$flux at the detector equals $\Phi_\tau = \Phi_\mu^0
P_{\mu\tau}$.  To be detected as a $\nu_\mu^{CC}-$ event in the same
muon energy bin the $\nu_\tau$ energy needs to be $\eta \sim 2.5$
times higher than a $\nu_\mu$ energy.  Thus, the rate of
$\nu_\mu^{CC}-$ events in the neutrino energy and zenith angle bin
$ij$ equals
\ba
N_{i,j} &=& 2\pi \int_{\Delta_i \cos\theta_z} d\cos\theta_z \int_{\Delta_jE} dE~ 
A_{\rm eff} (E, \theta_z)
\nonumber \\ & \times & 
\left[ \Phi_\mu^0 (E, \theta_z) P_{\mu\mu} (E, \theta_z) + 
 \Phi_\mu^0 (\eta E, \theta_z) \epsilon P_{\mu\tau} (\eta E, \theta_z)\right]
\nonumber \\ & + & ~{\rm antineutrinos}.  
\label{evts}
\ea
We use the DeepCore ``Filter'' effective area $A_{\rm eff}$ from
Ref. ~\cite{deepcore} and the atmospheric neutrino fluxes model of
Ref.~\cite{Honda:1995hz}.  The DeepCore effective area is largely
insensitive to the zenith angle~\cite{cowen} so that $A_{\rm
eff}(E, \theta_z) \approx A_{\rm eff}(E)$.  The contribution of the
$\tau-$decays to the total $\nu_\mu^{CC}-$event rate (the second term
in Eq. (\ref{evts})) is $\lesssim 5\%$. Since at low energies
secondary leptons from the $\nu N$ interactions are scattered at
larger angle from the directions of the primary neutrinos, we use
large zenith angle bins $\Delta_i(\cos\theta_z) = 0.2$, as compared to
the bins for $\gtrsim 100$ GeV data ~\cite{Abbasi:2010ie}.  We assume
that for $\nu_\mu^{CC}$ events the neutrino energy will be
reconstructed with factor of 2 accuracy which implies detection of
both muon and the associated (hadron) cascade. Therefore we use the
energy bins $\Delta_jE = (E \div 2E)$.

Figure~\ref{fig:evts2} shows the zenith angle distributions of
$\nu_\mu^{CC}-$events at DeepCore in different energy intervals.  The
histograms represent the event rates with $\nu_s$ and without $\nu_s$
mixing correspondingly for NH (left) and IH (right).  The rates
decrease with increase of $|\cos \theta_z|$ which reflects the dip in
the oscillation probability. The decrease is steeper in the low energy
range which covers the minimum of the dip.  With decrease of the
$\nu_s-$mixing the difference of histograms decreases as
$\sin^2\alpha$.  Due to summation of the neutrino and antineutrino
signals the total $\nu_s-$effect on the number of events is smaller
than the effect on probabilities.  Still the compensation is not
complete due to difference of the original fluxes and cross-sections
of the neutrinos and antineutrinos.  For IH the effect is smaller than
for NH, especially for deep (near vertical) trajectories. The reason
is that for IH relative deviation of the $\nu$ survival probability
from that without $\nu_s-$mixing is smaller than deviation for
$\bar{\nu}$ probability (see Fig. \ref{fig:prob_dms}).  This is
compensated by larger flux and cross section of $\nu$.  Notice that
difference of the event numbers in each of the zenith angle bins can
reach 2 - 3 $\sigma$ after 1 year exposure.

We define the suppression factor as the ratio of the rate $N_{i,j}$
with $\nu_s-$mixing to the same rate without $\nu_s$ mixing (see also
Ref.~\cite{Razzaque:2011ab}).  Figures~\ref{fig:suppressionNH}
and \ref{fig:suppressionIH} show the suppression factors as functions
of zenith angle in different $\nu_\mu$ energy ranges for NH and IH
correspondingly.  As can be seen from Fig. \ref{fig:suppressionNH} for
NH the $\nu_s$ mixing mostly affects the vertically upcoming events in
the lowest, 15--30 GeV, energy bin.  The excess of events in this bin
is due to enhanced ${\bar \nu}_\mu-$survival probability ${\bar
P}_{\mu\mu}$ near the oscillation minimum at $\sim 30$~GeV (see
Figs.~\ref{fig:prob} and \ref{fig:prob_mm}).  For IH
(Fig.~\ref{fig:suppressionIH}) the next bin $(-0.8 \div -0.6)$ rate is
equally affected.  The distributions depend on $\Delta m^2_{32}$
weakly.

In the bottom panels of Figures ~\ref{fig:suppressionNH}
and \ref{fig:suppressionIH} we show possible modification of the ratio
due to inprecise knowledge of $\Delta m^2_{32}$. This is illustrated
by the ratio of the two distributions without $\nu_s$ for two
different values of $\Delta m^2_{32}$: the ``true'' (observed) value
and the ``fit'' value. Variations of $\Delta m^2_{32}$ produce shift
of the dip and therefore qualitatively similar distortion of the
ratio. Quantitatively the effects are different: the effect of $\Delta
m^2_{32}$ variations is smaller than the one of $\nu_s$ at high
energies.  For IH and $|\cos \theta_z| > 0.4$ the two effects have
opposite signs, {\it etc}.. In analysis of data one should use $\Delta
m^2_{32}$ as fit parameter varying in the range allowed by accelerator
experiment measurements.  In future $\Delta m^2_{32}$ will be
determined with accuracy $10^{-4}$ eV$^2$ which will substantially
reduce the uncertainty.
 
\bfig
\includegraphics[trim=0.12in 0.37in 0.55in 1.in, clip, width=4.2in]{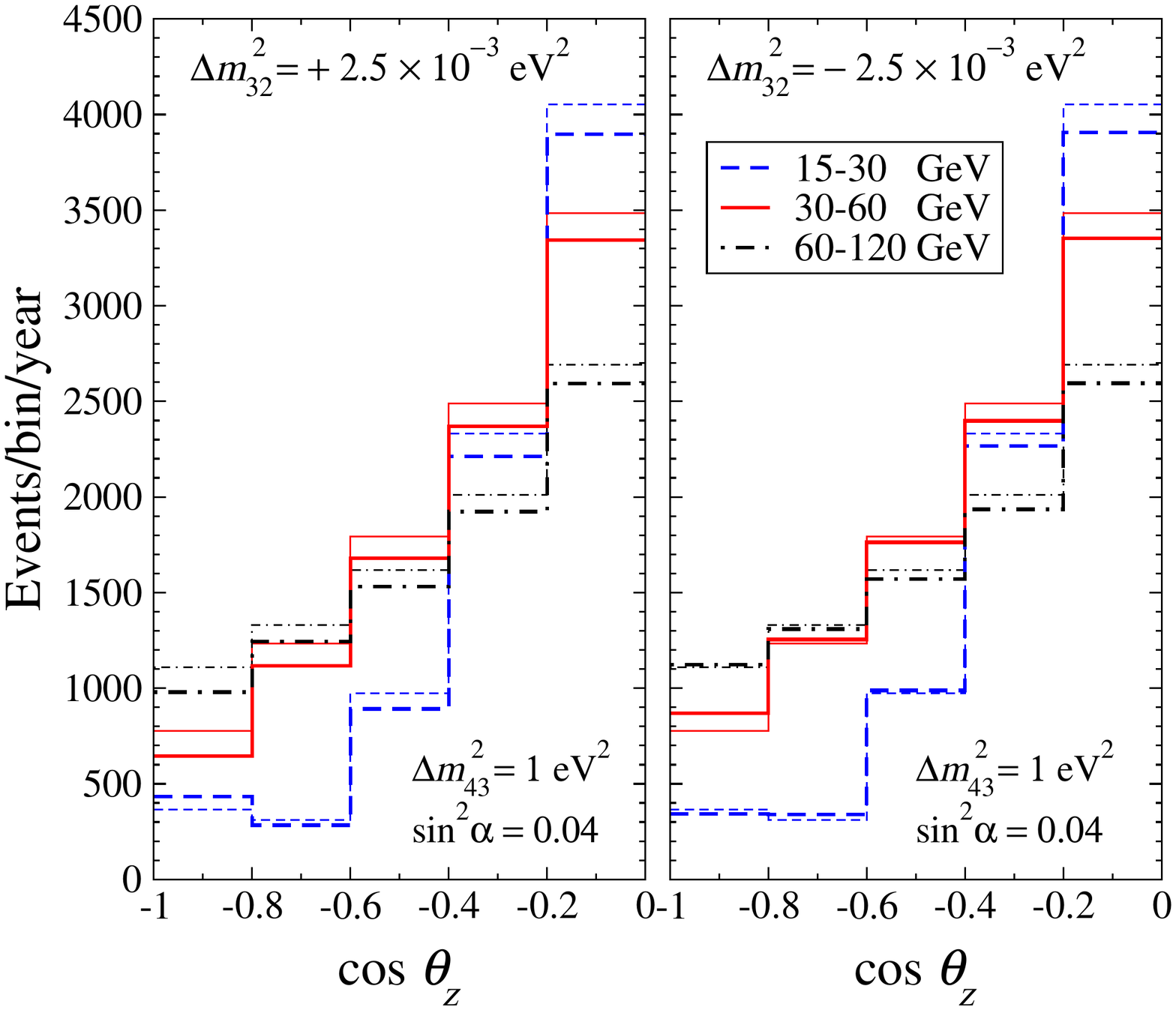}
\caption{
The $\nu_\mu^{CC}$ event rates in different energy and zenith angle
bins at IceCube DeepCore.  The histograms with thick lines correspond
to events with $\nu_s-$mixing while the histograms with thin lines
correspond to events without $\nu_s-$mixing. {\it Left panel}: normal
mass hierarchy, {\it right panel}: inverted mass hierarchy.
}
\label{fig:evts2}
\efig

\bfig
\includegraphics[trim=0.2in 0.45in 3.7in 1.1in, clip, width=3in]{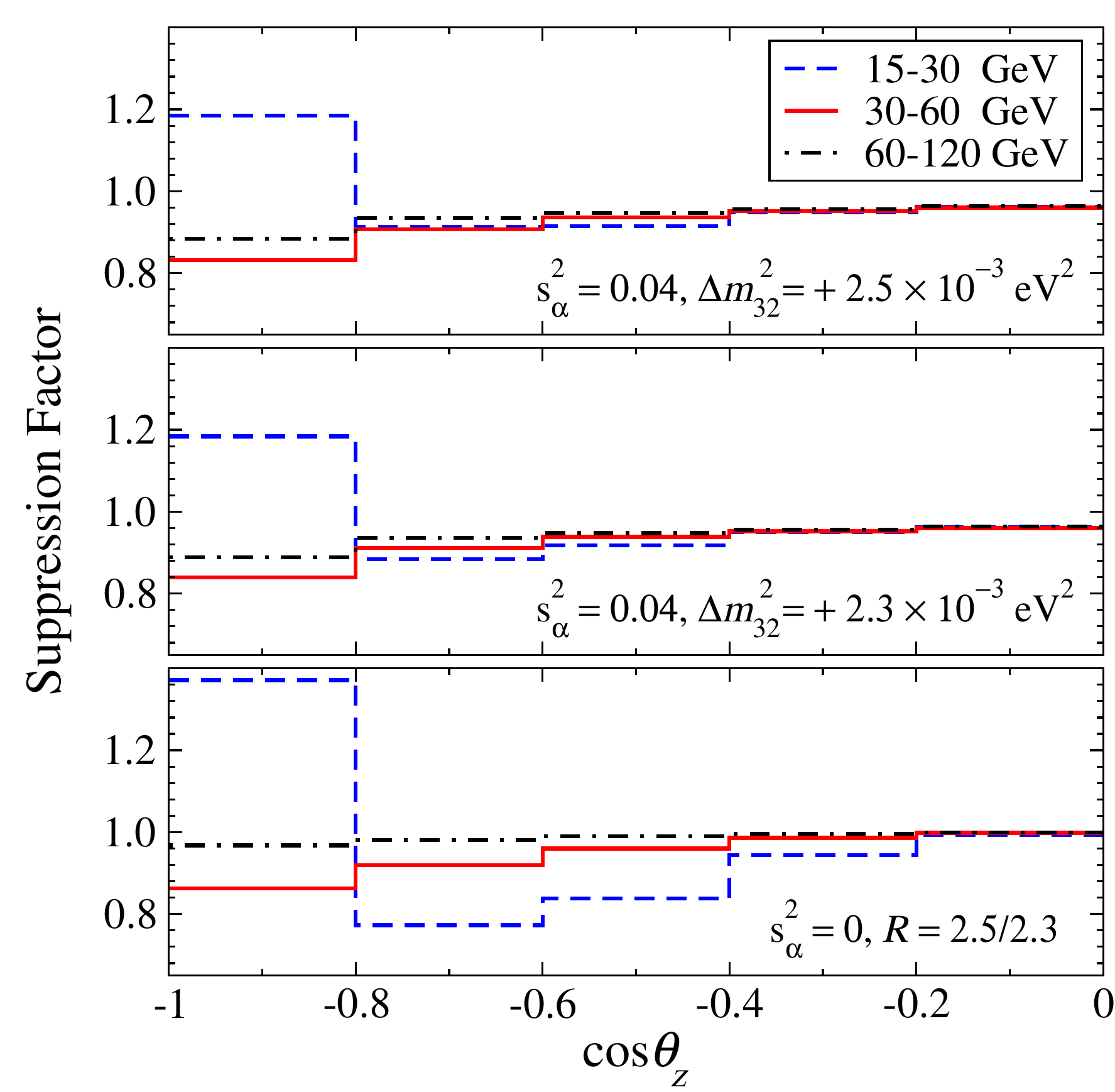}
\caption{
The suppression factor of $\nu_\mu^{CC}-$events as function of the
zenith angle for different energy intervals and two values of $\Delta
m^2_{32}$ (top and middle panels).  The bottom panel shows the ratio
of event rates for two different $\Delta m^2_{32}$ without sterile
neutrinos. The normal mass hierarchy is assumed.
}
\label{fig:suppressionNH}
\efig

\bfig
\includegraphics[trim=0.2in 0.45in 3.7in 1.1in, clip, width=3in]{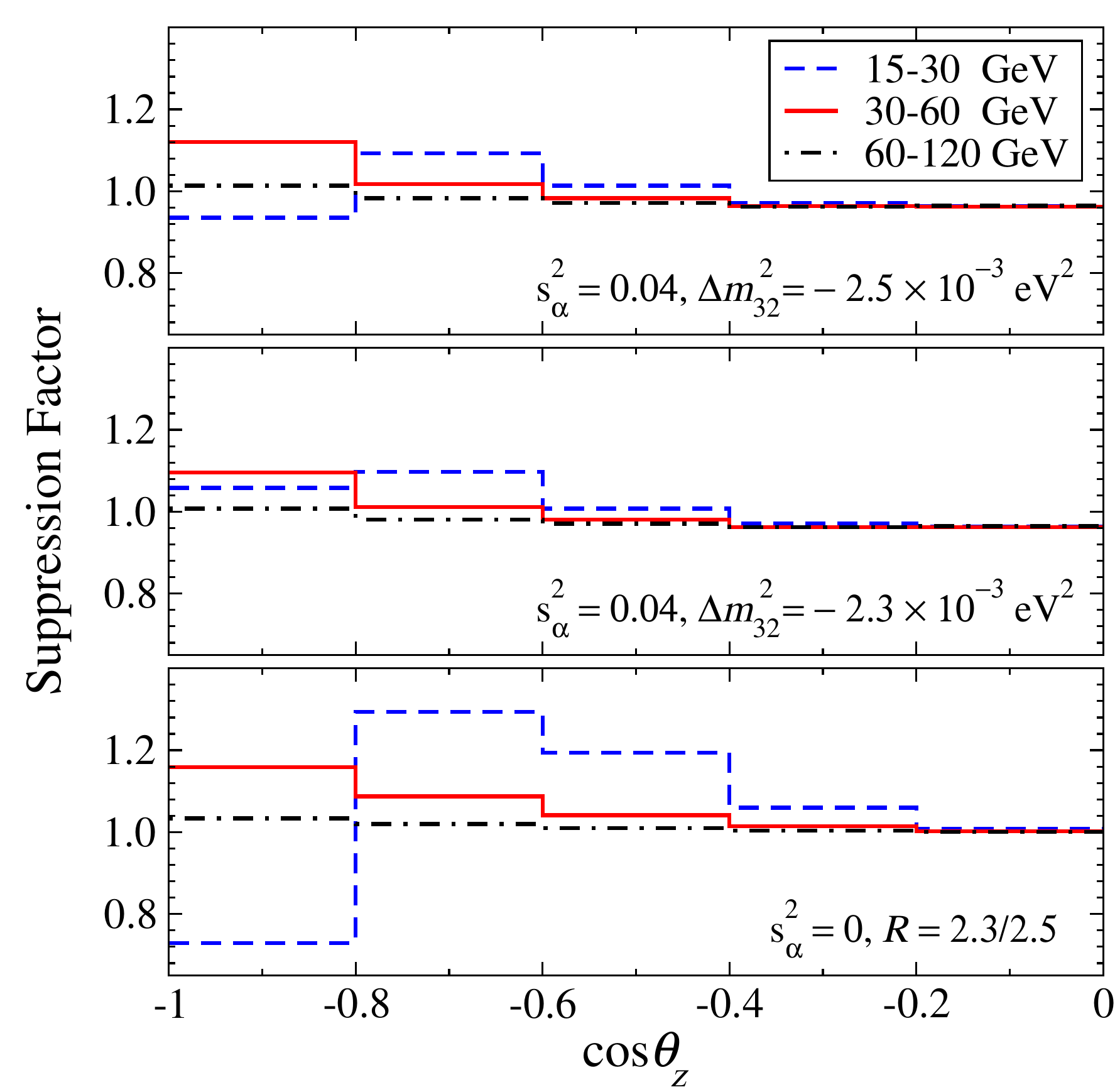}
\caption{
The same as in Fig.~\ref{fig:suppressionNH} for the inverted mass
hierarchy.
}
\label{fig:suppressionIH}
\efig

\section{Estimation of sensitivity to mixing}

\begin{table}
\caption{\label{tab1} 
Minimal value $\chi^2$ and the corresponding values of the
normalization $C$ and tilt parameter $\tau$ from the analysis of the
tentative energy and zenith angle distributions of DeepCore event
rates with and without $\nu_s$ mixing.  The values within brackets are
calculated using 5\% uncorrelated systematic uncertainties in addition
to the statistical and overall normalization uncertainties. The
analysis is performed for two different values of mixing angle
$\alpha$ and two exposure times.}
\begin{ruledtabular}
\begin{tabular}{ccccc}
$\sin^2\alpha$ & Exposure &  $\chi^2_{\rm min}$ & $C$ & $\tau$ \\ 
\hline
0.02     & 1 yr & 7.62 [3.47] & 1.03 [1.03] & $-0.03$ [$-0.02$] \\
 (NH)    & 5 yr & 38.1 [5.59] & 1.03 [1.03] & $-0.03$ [$-0.02$] \\
\hline
0.04     & 1 yr  & 41.1 [18.7] & 1.06 [1.06] & $-0.05$ [$-0.02$] \\
 (NH)    & 5 yr  & 205 [29.7]  & 1.06 [1.05] & $-0.05$ [$-0.00$]  \\
\hline
0.02     & 1 yr &  4.50 [1.84] &  1.00 [1.00]   & 0.04 [ 0.04] \\
 (IH)    & 5 yr &  22.5 [2.92] &  1.00 [1.00]   &  0.04 [ 0.04]\\
\hline
0.04     & 1 yr  & 14.8 [5.49] &  1.01 [1.01] &  0.09 [ 0.10]  \\
 (IH)    & 5 yr  & 73.8 [8.47] &  1.01 [1.00] &  0.09 [ 0.10] 
\end{tabular}
\end{ruledtabular}
\end{table}

To estimate sensitivity of the DeepCore to sterile neutrinos we
perform a simple $\chi^2$ analysis of the data generated in assumption
of zero $\nu_s$ mixing.  We assume that $\Delta m^2_{32}$ is known
precisely.  We fit these ``data''with number of events obtained in the
presence of $\nu_s-$mixing. To take into account possible systematic
errors, as in Ref.~\cite{Razzaque:2011ab}, we introduce an overall
normalization factor $C$ and a tilt parameter $\tau$ for the zenith
angle distribution of the event rates in Eq.~(\ref{evts}) as
\be
N_{i,j}^{\rm mod}(C, \tau; \sin^2\alpha) = C[1 + \tau (\cos\theta_i + 0.5)]
N_{i,j}(\sin^2\alpha).
\ee
We use $\Delta m_{43}^2 = 1$~eV$^2$. Defining the ``null''
distribution of zenith angle events without $\nu_s-$mixing as
$N_{i,j}^{\rm null} \equiv N_{i,j} (C=1,~\tau=0; ~\alpha = 0)$, we
calculate $\chi^2$ value as
\be
\chi^2 (C, \tau, \sin^2\alpha) = \sum_{i, j}
\frac{[N_{i,j}^{\rm null} - N_{i,j}^{\rm mod} 
(C, \tau, \sin^2\alpha)]^2}{N_{i,j}^{\rm null}}.
\ee
Then $\chi^2$ is minimized by varying $C$ and $\tau$.  The results,
$\chi^2_{min}$ and the corresponding values of $C$ and $\tau$ are
listed in the Table~\ref{tab1} for 1~yr and 5~yr of exposure and for
two different values of $\sin^2\alpha$.  According to
Fig.~\ref{fig:evts2}, already after 1 year of data taking the
statistical error in each bin is expected to be $\sim (2$--$4)\%$.
However, as pointed out in Ref.~\cite{Halzen:2011yq}, systematic
errors for IceCube are not well defined yet and will likely dominate
statistical errors.  To illustrate this effect on the sensitivity to
$\nu_s$ we performed $\chi^2$ analysis by introducing a $5\%$
uncorrelated systematic error for each bin in addition to the
statistical and correlated systematic (slope and normalization)
errors. The values of $\chi^2$ and ($C, \tau$) in this case are
reported in Table~\ref{tab1} within brackets.  The $\chi^2$ values in
the Table should be compared with $\chi^2_{\rm min} = 0$ which
corresponds to zero mixing with $\nu_s$.

For NH the obtained $\chi^2$ values with 2 degrees of freedom mean
that even with $5\%$ additional systematic uncertainties,
$\sin^2\alpha = 0.04$ or $|U_{\mu 4}|^2 = 0.02$ can be excluded at $>
99.9\%$ CL with one year of DeepCore data.  As a result, the
oscillation interpretation of LSND/MiniBooNE
data~\cite{lsnd,miniboone} that requires $|U_{\mu 4}|^2 \gtrsim 0.03$
for $\Delta m_{43}^2 \lesssim 1$~eV$^2$ will be severely constrained.

In the case of $\sin^2\alpha = 0.02$ or $|U_{\mu 4 }|^2 = 0.01$ five
years of DeepCore data would be required to reach $\sim 95\%$ CL for
exclusion.  This is comparable with the current bound from the MINOS
experiment~\cite{minos2011}.  A mixing angle smaller than
$\sin^2\alpha \sim 0.02$ cannot be excluded with DeepCore data alone.
For IH the effects is substantially smaller and only $|U_{\mu
4}|^2 \gtrsim 0.02$ can be tested.

We estimate that the present uncertainty in $\Delta m^2_{32}$ reduces
the sensitivity to $\nu_s-$mixing by factor of 2, {\it i.e.}  $|U_{\mu
4}|^2 = 0.02$ (for NH) can be excluded at $> 99.9\%$ CL after 4 years
of the data taking.

\section{Discussion and Conclusion} 

A combined analysis of the IceCube high-energy data which cover the
MSW resonance dip at TeV energies, together with DeepCore low energy
data will improve greatly the sensitivity to $\nu_s$ mixing
parameters.  Furthermore, a distinction between the flavor- and
mass-mixing schemes will be possible since DeepCore event
distributions are affected in the case of mass-mixing only, as we have
studied here, while the IceCube event distributions are affected in
both mixing schemes~\cite{Razzaque:2011ab}.

Notice that search for sterile neutrinos with IceCube and DeepCore are
complementary to that with MINOS and other accelerator experiments.
Indeed, IceCube measures the mass squared difference and mixing at
much higher energies and with large matter effect.  In scenarios with
energy and enviroment dependent neutrino masses, the mass and mixing
parameters may turn out to be substantially different in this two
setups.

We considered the oscillation effects for specific mixing scheme and
specific values of parameters and presented an illustrative analysis
of simulated data.  General analysis would include the sub-leading
effects of the non-zero 1-3 mixing, possible effects of the
CP-violation associated to existence of sterile neutrino,
consideration of different values of $|U_{\tau 4}|$, {\it etc.}

In conclusion, we have found that mixing of the eV scale mass $\nu_s$
indicated by LSND/MiniBooNE, substantially affects the oscillation
probabilities in the range 15 - 120 GeV accessible to IceCube DeepCore
experiment.  The effect is opposite in $\nu-$ and $\bar\nu-$ channels
and can reach $\sim O(1)$ size at the probability level. The $\nu_s$
mixing produces a shift and modifies shape of the oscillation dip. The
effect is different for the normal and inverted mass hierarchies of
the active neutrinos. Therefore, in principle, the hierarchy can be
identified, if the sterile neutrinos with relatively large mixing
exist.  The zenith angle distributions of the $\nu_\mu^{CC}$ events in
different energy ranges are changed.  However, due to summation of
$\nu-$ and $\bar\nu-$ signals the effect of $\nu_s-$mixing and
oscillations on the number of events is much weaker. Uncertainty in
$\Delta m_{32}^2$ reduces sensitivity of DeepCore to sterile
neutrinos.  Nevertheless we find that Deep Core can probe the $\nu_s$
mixing down to $U_{\mu 4} = 0.01 - 0.02$ for wide range of $\Delta
m_{43}^2$.

\section*{Acknowledgments} 

We are grateful to D.~F.~Cowen and J.~Koskinen for correspondence on
the design and performance of the IceCube DeepCore.  Work of S.R. was
funded by NASA, and performed at and while under contract with the
U.S.~Naval Research Laboratory.  S.R.  would also like to thank the
Abdus Salam ICTP for hospitality while completing the final
manuscript.

\end{document}